\documentclass[prd,aps,a4paper,nofootinbib,eqsecnum,preprint]{revtex4}  

\usepackage{mathrsfs}
\usepackage{multirow}

\newif\ifusesec
\usesectrue  
   
\usepackage{graphicx} 
\usepackage{amsmath,amsfonts,amssymb}
\usepackage{mathtools}
\usepackage{color}

\newcommand{\beq}{\begin{equation}}
\newcommand{\eeq}{\end{equation}}
\newcommand{\bea}{\begin{eqnarray}}
\newcommand{\eea}{\end{eqnarray}}

\begin{document}

\title{Bekenstein-Hawking temperature from the Schwarzian}

\author{Donato Bini$^{1}$, Giampiero Esposito$^{2,3}$}

\affiliation{$^1$Istituto per le Applicazioni del Calcolo ``M. Picone'', 
CNR, I-00185 Rome, Italy\\
%Orcid: 0000-0002-5237-769X}
}

\affiliation{$^2$Dipartimento di Fisica ``Ettore Pancini'', \\
Complesso Universitario di Monte S. Angelo,
Via Cintia Edificio 6, 80126 Napoli, Italy\\
%Orcid: 0000-0001-5930-8366
} 

\affiliation{$^3$Istituto Nazionale di Fisica Nucleare, Sezione di Napoli, \\
Complesso Universitario di Monte S. Angelo,
Via Cintia Edificio 6, 80126 Napoli, Italy}

\begin{abstract}
Hawking's original derivation of particle creation by black holes 
in Schwarzschild spacetime exploits, among various concepts, the
exponential dependence on the retarded time variable $u$ 
of the affine parameter $\lambda$ of the null geodesics that
are integral curves of the null vector field orthogonal to the Killing horizon.
This exponential law implies that the Schwarzian derivative 
of $\lambda$ with respect to $u$ is minus a half the square of surface gravity. 
The black hole Killing horizon inherits an intrinsic projective structure, 
and the squared surface gravity is the invariant characterizing such a structure.

There is therefore evidence that 
the Bekenstein-Hawking temperature is completely determined from the projective
structure on the Killing horizon. As a further test, it is here shown that,
in a spacetime model with variable mass parameter, the logarithmic derivative of 
surface gravity is determined by the Schwarzian of the affine parameter.
The Schwarzian in Schwarzschild and Kerr geometries is also studied in detail. 
All these properties are a first step towards proving that black hole
thermodynamics finds its mathematical foundations in the projective geometry
of Killing horizons. Such a research program can be applied to
the power radiated from a black hole, the rate of change of the black hole
mass with respect to the area of the event horizon, the fundamental imaginary frequency 
of quasinormal modes (and hence the decay rate of black hole perturbations).

\end{abstract}

\date{\today}
\maketitle

\section{Introduction}

Ever since general relativity was developed by Einstein \cite{E1}, many languages have  
been created by the scientific community in order to achieve a geometric formulation
of all phenomena and all interactions. The program that we initiate here is part
of this broader framework, with emphasis on the geometric interplay between 
gravitational and statistical physics. In particular, 
since Hawking discovered the effect of particle creation by black 
holes \cite{H1,H2}, many efforts
have been devoted to studying its relevance for the foundations of general relativity and
quantum gravity, and the modern approaches to quantum gravity have to pass the testbed 
of black hole thermodynamics \cite{B1,B2,B3,H3,Witten} 
and Hawking's particle creation. However, bearing in 
mind that this is an effect of quantum field theory in a curved spacetime that is still
classical, it remains of interest to devote further attention to the classical aspects
of the appropriate geometric framework. For this purpose, we here summarize the basic
concepts that we need. 

In our work we focus on the affine parameter $\lambda$ (see now Appendix A for 
technical details) for the null geodesics that are integral curves of the (null)
vector field $l$ orthogonal to the timelike Killing vector field $K$ of Schwarzschild 
spacetime (which becomes null on the horizon).
Such an affine parameter is a smooth function of the retarded time variable $u$ on the
past horizon\footnote{This is the nomenclature appropriate for the particle creation
process studied by Hawking.}:
\begin{equation}
u=t-r_*=t-r-2M \log \left | \frac{r}{2M}-1 \right| ,
\label{(1.1)}
\end{equation}
where $r_*=r+2M \log \left | \frac{r}{2M}-1 \right|$ denotes the tortoise coordinate
of Schwarzschild spacetime.
The surface gravity $\kappa$ can be defined from the differential equation 
\begin{equation}
\nabla_{K}K = -\kappa \; K,
\label{(1.2)}
\end{equation}
and the affine parameter $\lambda$ is found to obey the exponential law
(see Ref. \cite{H2} and our Appendix)
\begin{equation}
\lambda=-C \; e^{-\kappa u},
\label{(1.3)}
\end{equation}
where $C$ is a constant.

Section 2 describes some basic properties of a fundamental concept in projective 
geometry, i.e., the Schwarzian of a smooth function, including its
relation with the cross-ratio. Section 3 evaluates the Schwarzian of the affine 
parameter on the black hole horizon in a static model, 
and studies its meaning from the point of view of projective geometry. 
Section 4 studies the Schwarzian in black hole spacetimes, while
Sec. 5 obtains the relation between Schwarzian and
surface gravity in a spacetime model where the mass depends
on the retarded time variable. 
Section 6 lists further consequences for black hole physics,
while some relevant open problems are mentioned in Sec. 7.
Important background material is presented in the Appendices.

\section{Some basic properties of the Schwarzian}

In the twentieth century, projective differential geometry 
(Appendix B) was developed 
by the American \cite{Wilc,Thomas,Lane1,Lane2}, Italian 
\cite{F1,F2,F3,F4,T1,T2,Bompiani,Segre,Villa} 
and French Schools \cite{C1,C2},
but the modern techniques differ sharply from 
the original formulations. For a detailed synthesis of a modern 
perspective, the reader may be 
referred to Refs. \cite{Ferapontov,Ov1,Eastwood,Ov2}. 

For our purposes, we have to consider the Schwarzian $S$ of differentiable 
functions\footnote{Indeed, Ref. \cite{Ov1} defines the action of $S$ on
diffeomorphisms of the real projective line, but its action can be defined
on smooth functions as well.} $f$ of
a real variable $x$, following Ref. \cite{Ov2}:
\begin{equation}
S(f(x))=\frac{f'''(x)}{f'(x)}-\frac{3}{2}
\left(\frac{f''(x)}{f'(x)}\right)^{2}.
\label{(2.1)}
\end{equation}
This nonlinear differential operator can be obtained from the nonlinear operator in the
Riccati equation \cite{Ferapontov}:
$$
T(\eta(x))=\eta'(x)-\frac{1}{2}\eta^{2}(x),
$$
upon expressing $\eta$ as the logarithmic derivative of $f'$: 
$\eta=f''/f'$. A first important
property of the map $S$ is that the necessary and sufficient condition for
two functions $f$ and $h$ to have the same Schwarzian: $S(f)=S(h)$ is 
(hereafter $a,b,c,d$ are real or complex numbers)
\begin{equation}
h(x)=\frac{(af(x)+b)}{(cf(x)+d)}, \; ad-bc \not =0.
\label{(2.2)}
\end{equation}
In particular, $S(f)$ vanishes if and only if $f$ is a fractional linear transformation, i.e.
\begin{equation}
S(f)=0 \Longleftrightarrow f(x)=\frac{(ax+b)}{(cx+d)}.
\label{(2.3)}
\end{equation}
By convention, the value of such a map $f$ at $-\frac{d}{c}$ is $\infty$, and the
value of $f$ at $\infty$ is $\frac{a}{c}$. Fractional linear maps can be viewed as
diffeomorphisms of ${\mathbb R}{\mathbb P}^{1}$. The transformations 
\eqref{(2.3)} with real coefficients 
form the group of projective symmetries of ${\mathbb R}{\mathbb P}^{1}$:
\begin{equation}
PSL(2,{\mathbb R})=SL(2,{\mathbb R})/\delta,
\label{(2.4)}
\end{equation}
where $\delta$ is the homeomorphism defined by 
\begin{equation}
\delta(a,b,c,d)=(-a,-b,-c,-d).
\label{(2.5)}
\end{equation}

Given two diffeomorphisms $f,h$ of ${\mathbb R}{\mathbb P}^{1}$, the Schwarzian of 
their composition $h \odot f$ satisfies the condition
\begin{equation}
S(h \odot f)=S(h) \odot f + S(f),
\label{(2.6)}
\end{equation}
with the understanding that 
\begin{equation}
(u \odot f)(x)=u(f(x))(f'(x))^{2}.
\label{(2.7)}
\end{equation}

\subsection{Schwarzian and cross-ratio}

Given a quadruple of points in ${\mathbb P}^{1}$, one can choose an affine coordinate
that represents the points by four numbers $t_{1},t_{2},t_{3},t_{4}$; the {\it cross-ratio}
\begin{equation}
[t_{1},t_{2},t_{3},t_{4}]=\frac{(t_{1}-t_{3})(t_{2}-t_{4})}{(t_{1}-t_{2})(t_{3}-t_{4})}
\label{(2.8)}
\end{equation}
is invariant under the projective transformations of the projective line. Interestingly,
this discrete invariant is related to the Schwarzian, because, given a diffeomorphism 
$f$ of ${\mathbb R}{\mathbb P}^{1}$, the Schwarzian measures how $f$ affects the 
cross-ratio of infinitesimally close points. Indeed, let $t$ be a point in
${\mathbb R}{\mathbb P}^{1}$ and let $\nu$ be a tangent vector to 
${\mathbb R}{\mathbb P}^{1}$ at $t$. Such a vector $\nu$ can be extended to a vector
field in a neighbourhood of $t$, with corresponding local flow denoted by $\phi_{s}$.
Consider now the four points $t, t_{1}=\phi_{\varepsilon}(t), 
t_{2}=\phi_{2 \varepsilon}(t), t_{3}=\phi_{3 \varepsilon}(t)$. The cross-ratio does
not change to first order in $\varepsilon$, because
\begin{equation}
[f(t),f(t_{1}),f(t_{2}),f(t_{3})]=[t,t_{1},t_{2},t_{3}]
-2 \varepsilon^{2}S(f)(t)+{\rm O}(\varepsilon^{3}).
\label{(2.9)}
\end{equation}
The coefficient of the term quadratic in $\varepsilon$ depends on the diffeomorphism
$f$, the point $t$, the tangent vector $\nu$, but not on the extension of $\nu$ to a
vector field. It is homogeneous of degree $2$ in $\nu$.
In the early days of projective differential geometry, it was an important result that 
the Fubini projective line element of a surface is a suitable cross-ratio 
(Ref. \cite{T1} and our Appendix B). 

\subsection{Action of $S$ on elementary functions}

When acting on $f(x)=ax^n$ the Schwarzian leads to a result proportional to $x^{-2}$
\begin{equation}
S(ax^n)=\frac{(1-n^2)}{2x^2}=S \left(\frac{1}{ax^n}\right).
\end{equation}
Therefore, apart from the trivial case $n=\pm 1$ for which the Schwarzian 
vanishes, the case $n=2$ is interesting in the sense that one has
\begin{equation}
S \left(\frac{1}{ax^2}\right)=-\frac{3}{2x^2}
\end{equation}
which identifies the \lq\lq fixed function" $f(x)=-\frac{3}{2x^2}$.
The same result occurs when $f(x)=\log(x)$,
\begin{equation}
S(\log(x))=\frac{1}{2x^2}
\end{equation}
whereas
\begin{equation}
S(e^{x})=-\frac12,\qquad S(\tan(x))=2,
\end{equation}
and
\begin{equation}
S(\sin(x))=\frac12\frac{(\cos^2(x)-3)}{\cos^2(x)} ,\qquad 
S(\cos(x))=-\frac12\frac{(\cos^2(x)+2)}{\sin^2(x)} .
\end{equation}
Moreover, for example
\begin{equation}
S \left(\sum_{k=0}^\infty x^{k} \right)=S \left(\frac{1}{(1-x)}\right)=S(1-x)=0,
\end{equation}
and
\begin{equation}
S({\rm arctan}(x))=-\frac{2}{(1+x^2)^2}.
\end{equation}
The following properties also holds
\begin{equation}
S(c+f(x))=S(f(x)),
\end{equation}
and
\begin{equation}
S(f(x)^n)=S(f(x))+\frac{(1-n^2)}{2 f^2(x)}\left( f'(x)\right)^{2}.
\end{equation}

\subsection{Iterated Schwarzian derivatives}

We have seen that for a function $f$ of a single variable $t$
\begin{equation}
S(f)(t)=0 \Longleftrightarrow f(t)=\frac{(at+b)}{(ct+d)}
\label{(2.10)}
\end{equation}
where 
\begin{equation}
S(f)(t)=\frac{f^{(3)}}{f'}-\frac{3}{2}
\left(\frac{f''}{f'}\right)^{2},
\label{(2.11)}
\end{equation}
i.e., the kernel of the Schwarzian consists of fractional linear transformations. 
Let us note that we often do not explicitly display the argument and simply write 
$S(f)$ instead of $S(f)(t)$ or $S(f(t))$, for simplicity.
This is called a first degree Schwarzian derivative.
Fractional quadratic functions, which are of the type 
\begin{equation}
f=\frac{(at^2+bt +c)}{(dt^2+et+f)},
\label{(2.12)}
\end{equation}
form instead the kernel of a more complicated operator $S_2$.
Other variations of the Schwarzian derivative are interesting, that is 
its iteration and looking for the existence of fixed points.
In other words, one may study the result of multiple iterations of $S$
\begin{equation}
S(S(S(\ldots f)))(x)=0
\end{equation}
as well as the special class of functions satisfying the following fixed point relation
\begin{equation}
S(f)(x)=c f(x)
\end{equation}
which both lead to nonlinear ordinary differential equations.
We will discuss these cases below, within explicit examples.

\subsection{Schwarzian and curvature}

Let us consider the Lorentz 2-plane, with metric $g=\frac{1}{2}(dx \otimes dy 
+ dy \otimes dx)$ (signature $-+$), and
let $f$ be a spatial curve with image $f(x)$, 
with spatial unit tangent vector given by
\begin{equation}
U=\frac{1}{\sqrt{f'(x)}}
\left(\frac{\partial}{\partial x}+f'(x)\frac{\partial}{\partial y}\right),
\qquad U\cdot U=g(U,U)=+1.
\end{equation}
The corresponding (timelike) acceleration vector of $U^\alpha$ is  
$a(U)^\alpha=\nabla_U U^\alpha$ and is given by
\begin{equation}
a(U)=\nabla_U U=\frac12 \frac{f''(x)}{f'(x)}
\left(-\frac{1}{f'(x)}\frac{\partial}{\partial x} 
+\frac{\partial}{\partial y} \right).
\end{equation}
Its squared magnitude reads as
\begin{equation}
a(U)\cdot a(U)=g(a(U),a(U))=-\frac14  \frac{(f''(x))^2}{(f'(x))^3},
\end{equation}
and hence the curvature \cite{Ov1,Ov2} is
\begin{equation}
\kappa(x) \equiv \sqrt{-a(U)\cdot a(U)}=\frac12 \frac{f''(x)}{(f'(x))^{3/2}}.
\end{equation}
The following relation holds:
\begin{equation}
\label{(2.14)}
\kappa'(x) =\frac12 \frac{S(f(x))}{\sqrt{f'(x)}}.
\end{equation}
What is truly interesting in this example is that Eq. \eqref{(2.14)}
is a first indication that the Schwarzian derivative can be related
to a curvature.

\subsection{Schwarzian and metric}

We have seen that the application of the Schwarzian operation to a polynomial function 
generates a \lq\lq simpler" polynomial one, e.g.,
\begin{equation}
S(x^n)=\frac{(1-n^2)}{2x^2}.
\end{equation}
In other words, every power of $x$ (except for the case $n=\pm 1$) is mapped into $x^{-2}$.
Let us consider a spacetime with metric depending on a single variable, say $t$
\begin{equation}
g=g_{\alpha \beta}(t)dx^\alpha \otimes dx^\beta,
\end{equation}
where $x^\alpha=(t,x,y,z)$.
One can then consider the action of the Schwarzian on $g_{\alpha\beta}$ as defining 
a new metric (which however is no longer a solution of Einstein's equations)
\begin{equation}
{\bar g}_{\alpha \beta}(t)=S(g_{\alpha \beta})(t)
\end{equation}
such that
\begin{equation}
{\bar g}={\bar g}_{\alpha \beta}(t)dx^\alpha \otimes dx^\beta.
\end{equation}
The question is: what geometrical properties are inherited by $\bar g_{\alpha \beta}(t)$ 
which where characterizing the original metric $g_{\alpha \beta}(t)$.

Let us answer this question with an explicit example.
For this purpose we consider the Kasner metric
\begin{equation}
g_{K}=-dt \otimes dt+t^{2p_1}dx \otimes dx+t^{2p_2}dy \otimes dy
+t^{2p_3}dz \otimes dz,
\end{equation}
with
\begin{equation}
\sum_{k=1}^{3}p_{k}=1=\sum_{k=1}^{3}(p_{k})^{2},
\end{equation}
i.e.,
\begin{equation}
g_{\alpha\beta}={\rm diag}[-1,t^{2p_1},t^{2p_2},t^{2p_3}].
\end{equation}
The Schwarzian-transformed metric is singular in 
four dimensions, in the sense that the 
$tt$ component of the metric vanishes, and the metric reduces to
\begin{equation}
{}^{(3)}g=\frac{1}{2t^2}[(1-4(p_1)^2)dx \otimes dx
+(1-4(p_2)^2)dy \otimes dy+(1-4(p_3)^2)dz \otimes dz].
\end{equation}
Apart from the $t$-dependent overall factor (one can assume for example 
$t=\frac{1}{\sqrt{2}}$), one deals actually with
\begin{equation}
{}^{(3)}g= (1-4(p_1)^2)dx \otimes dx
+(1-4(p_2)^2)dy \otimes dy+(1-4(p_3)^2)dz \otimes dz ,
\end{equation}
the new 3-metric is flat so that the corresponding geodesics are straight lines:
\begin{equation}
(1-4(p_1)^2)^{1/2}x=v_1  \lambda ,\ (1-4(p_2)^2)^{1/2}y=v_2  \lambda , 
(1-4(p_3)^2)^{1/2}z=v_3  \lambda .
\end{equation}
By contrast, timelike (for example) geodesics of the Kasner original metric 
are characterized by the 4-velocity
\begin{equation}
U=-\left(1+\frac{(p_1)^2}{t^{2p_1}} +\frac{(p_2)^2}{t^{2p_2}} 
+\frac{(p_3)^2}{t^{2p_3}} \right)^{1/2} dt +p_1 dx+p_2dy +p_3dz ,
\end{equation}
and only on the $t$=constant hypersurfaces do they reduce to straight lines.

Another simple application concerns the familiar Schwarzschild metric. 
If one considers its two-dimensional reduction by means of the hypersurfaces 
$\theta$=constant and $\phi$=constant, the metric becomes
\begin{equation}
g_{(t,r)}=-\left(1-\frac{2M}{r}\right)dt \otimes dt
+\frac{dr \otimes dr}{\left(1-\frac{2M}{r}\right)}.
\end{equation}
The two metric coefficients $g_{tt}$ and $g_{rr}$ are both fractional linear 
functions of the variable $r$ and hence they yield $0$ when acted upon by the Schwarzian 
(with respect to $r$): $S(g_{tt},r)=0=S(g_{rr},r)$.  
The application of the Schwarzian (with respect to $r$) to the complete 
Schwarzschild metric picks out therefore its angular part
\begin{equation}
S(g_{\theta\theta},r)=-\frac{3}{2r^4}g_{\theta\theta},\qquad S(g_{\phi\phi},r)=-\frac{3 }{2r^4}g_{\phi\phi},
\end{equation}
since the $t-r$ part is annihilated from it.

These types of calculations are original, but we are well aware that
the metric is not the natural object for the application of the Schwarzian. One has instead to look
at the projective structure resulting from the class of affine parameters, as
we do in the following section.

\section{Schwarzian of the affine parameter and projective structure of the Killing horizon}

By virtue of Eqs. \eqref{(1.3)} and \eqref{(2.1)} we find that
(the prime denoting differentiation with respect to the affine parameter $u$)
\begin{equation}
\lambda'(u)=\kappa C e^{-\kappa u}, \;
\lambda''(u)=-\kappa \lambda'(u), \;
\lambda'''(u)=\kappa^{2}\lambda'(u),
\label{(3.1)}
\end{equation}
and hence
\begin{equation}
S(\lambda(u))=\kappa^{2}-\frac{3}{2}\kappa^{2}
=-\frac{\kappa^{2}}{2}.
\label{(3.2)}
\end{equation}
This yields in turn our basic formula for the surface gravity of a
Schwarzschild black hole (of course, its actual value $\frac{1}{4M}$ in 
$G=c=1$ units is well known):
\begin{equation}
\kappa^{2}=-2 S(\lambda(u))=2 |S(\lambda(u))| 
\Longrightarrow \kappa = \sqrt{2} \sqrt{|S(\lambda(u))|}.
\label{(3.3)}
\end{equation}
As far as we can see, in light of the review material in Sec. 2 and our Eq.
\eqref{(3.3)}, the following remarks are now in order.
\vskip 0.3cm
\noindent
(i) On writing Eq. \eqref{(3.3)} 
we are saying that the surface gravity describes 
by how much the natural parametrization of the horizon generators differs from a
projective transformation, and $\kappa$ can be seen as the projective curvature
of the beam of horizon's generators (since the Schwarzian is projective curvature,
as shown for example in Ref. \cite{Ov1}).
\vskip 0.3cm
\noindent
(ii) The horizon inherits an intrinsic projective structure, and the squared surface gravity $\kappa$
is the invariant characterizing such a structure. Moreover, the physics at the horizon
does not depend on reparametrizations of the affine parameter which preserve the
cross-ratio. In other words, what affects $\kappa$ is precisely what affects the cross-ratio
along the generators, and $\kappa$ is what remains after taking out all transformations
which preserve the cross-ratio. Thus, surface gravity is a physical invariant and not a
coordinate artifact.
\vskip 0.3cm
\noindent
(iii) Since the Bekenstein-Hawking temperature is expressed by the formula
\begin{equation}
T_{BH}=\frac{\kappa}{2\pi},
\label{(3.4)}
\end{equation}
we find from Eqs. \eqref{(3.3)} and \eqref{(3.4)} that 
\begin{equation}
T_{BH}=\frac{1}{\pi \sqrt{2}} \sqrt{|S(\lambda(u))|}.
\label{(3.5)}
\end{equation}
Thus, the foundations of black-hole temperature lie in the projective geometry of the 
Killing horizon.
\vskip 0.3cm
\noindent 
(iv) We are not trading the spacetime manifold for a projective space, but we are
realizing that the equivalence class of affine parameters along the generators
of the horizon defines a natural projective structure on the horizon itself. Such
a projective structure is independent of coordinates, is preserved by Killing
symmetries and determines the surface gravity.

\subsection{Projective structure on the Killing horizon}

At the risk of slight repetitions, let us consider again the geometric framework of interest.
Each Killing horizon is generated by null curves that are null geodesics for the
Levi-Civita connection. This affine parameter $\lambda$ is not unique, because 
the linear combination ${\tilde \lambda}=a \lambda +b$ is an affine parameter as well. 
This is precisely the definition of projective structure on a curve. The equivalence 
class $[\lambda]$ of such affine parameters has three properties:
\vskip 0.3cm
\noindent
(1) independence of coordinates (since $\lambda$ depends only on the connection);
\vskip 0.3cm
\noindent
(2) preservation under Killing symmetry: if $\phi_{t}$ is the flow of $K$, it then
maps generators into generators and affine parameters into affine parameters;
\vskip 0.3cm
\noindent
(3) intrinsic nature at the horizon: $[\lambda]$ does not depend on how the horizon
is embedded into spacetime, but only on the Killing horizon structure. 

Thus, each horizon generator carries naturally a projective structure, i.e., a class
of affine parameters defined up to linear transformations here denoted by $T$:
\begin{equation}
T(\lambda)=a \lambda +b,
\label{(3.6)}
\end{equation}
with the associated scaling laws for Eq. \eqref{(1.2)}:
\begin{equation}
T(K)=\frac{1}{a} K, \; T(\kappa)=\frac{1}{a}\kappa.
\label{(3.7)}
\end{equation}
The surface gravity $\kappa$ is therefore defined up to a projective transformation, 
and the projective structure determines which transformations
are admissible. Equations \eqref{(3.7)} tells us that surface gravity changes in 
a controlled way but is not a projective invariant. As we said before, surface
gravity is precisely what cannot be eliminated after removing all transformations
that preserve the cross-ratio. 

\section{Schwarzian in black hole spacetimes}

Black hole spacetimes, like Schwarzschild and Kerr, have been 
extensively studied in general relativity. 
Let us consider the family of timelike circular equatorial orbits in these 
spacetimes (we assume the metric signature to be Lorentzian), including 
either geodesic or non-geodesic (accelerated) orbits.
The aim of this analysis is to make contact with the Schwarzian again when 
looking at the acceleration (curvature) of these special orbits.

Let us write their four-velocity $U$ as
\begin{equation}
U=\gamma(U,u)[u +\nu(U,u) e(u)_{\hat \phi}],\qquad 
\gamma(U,u)=(1-\nu^{2}(U,u))^{-1/2},
\end{equation}
where $u$ is any family of observers, e.g., static 
or zero angular momentum observers,
and $e(u)_{\hat \phi}$ denotes the unit vector of the azimuthal direction 
in the Local-Rest-Space of $u$, $LRS_u$ \cite{Jantzen:1992rg,Bini:1997ea,Bini:1997eb}. 
Here $\nu=\nu(U,u)$ (depending both on the orbit itself and on the observer 
who \lq\lq measures" it) is a single parameter used to label the family of 
circular orbits~\footnote{Even if the notation may appear heavy at the beginning 
then one immediately realizes how powerful is it.}.
In terms of coordinates one would write
\begin{equation}
U=\Gamma \left(\frac{\partial}{\partial t}
+\zeta \frac{\partial}{\partial \phi} \right),
\end{equation}
where now the parameter is the angular velocity $\zeta$ instead of the relative 
velocity with respect to the observers $u$. For example, if 
$u=u_{\rm stat}$ denotes the static family of observes the two parameters 
$\nu(U,u_{\rm stat})$ and $\zeta$ are simply related as
\begin{equation}
\nu(U,u_{\rm stat})=\frac{\sqrt{\gamma_{\phi\phi}}\zeta}{M(1-M_\phi\zeta)} ,
\end{equation}
having used the notation
\begin{equation}
ds^2=-M^2(dt-M_\phi d\phi)^2+\gamma_{\phi\phi}d\phi^2 +\gamma_{rr}dr^2
\end{equation}
for the metric adapted to the static obsevers in the equatorial plane.
Conversely, if $u=u_{\rm ZAMO}$ denotes the Zero Angular Momemtum (ZAMO) 
family of observers, the two parameters $\nu(U,u_{\rm stat})$ and $\zeta$ are simply related as
\begin{equation}
\nu(U,u_{\rm ZAMO})=\frac{\sqrt{g_{\phi\phi}}(\zeta+N^\phi)}{N} ,
\end{equation}
having used the notation
\begin{equation}
ds^2=-N^2dt^2+g_{\phi\phi}(d\phi+N^\phi dt)^2 +g_{rr}dr^2
\end{equation}
for the metric adapted to the ZAMOs in the equatorial plane.
$\Gamma$ is instead a normalization 
factor to ensure the timelike condition $U\cdot U=-1$.

By virtue of spacetime symmetries (i.e., reflection symmetry with 
respect to the equatorial plane), the four-acceleration of such orbits is purely 
radial, and hence is characterized by its magnitude $\kappa$ only.
In the Schwarzschild case (with static observers coinciding with 
zero angular momentum observers) the latter curvature reads
\begin{equation}
\label{k_schw}
\kappa(\nu) =k_S \frac{(\nu^2-\nu_g^2)}{(1-\nu^2)},
\end{equation}
where $k_S$ does not depend on $\nu$ (but depends on $r$) 
\begin{equation}
k_S= -\frac{1}{r}\sqrt{1-\frac{2M}{r}},
\end{equation}
and $\nu_g$ denotes the magnitude of the relative velocity of circular geodesics,
We assume that 
$$
u=\frac{1}{\sqrt{1-\frac{2M}{r}}}\frac{\partial}{\partial t}
$$ 
can be identified with 
the static family of observers (i.e., at rest with respect to the Schwarzschild coordinates)
\begin{equation}
\nu_g=\pm \frac{\sqrt{\frac{M}{r}}}{\sqrt{1-\frac{2M}{r}}},
\end{equation}
with the $\pm$ signs corresponding to co-rotating and counter-rotating circular 
geodesics with respect to the (reference, positive) counter-clockwise 
rotation of polar coordinates.

The two circular geodesics become null at $r=3M$ where
$|\nu_g| = 1$, and then spacelike for smaller radii.
At this radius $r=3M$ the curvature $\kappa(\nu)$ reduces to
$\kappa(\nu)= -k_S > 0$ which is independent of
the velocity $\nu$ of the test particle. For $2M < r < 3M$  the circular geodesics 
are both spacelike and $\kappa(\nu)$ is positive.

In Eq. \eqref{k_schw}, the function $\kappa(\nu)$ is a fractional linear relation 
in $\nu^2$ and hence its Schwarzian (taken with respect to the variable $\nu^2$ 
and not $\nu$) vanishes
\begin{equation}
S(\kappa(\nu),\nu^2)=0,
\end{equation}
where the indication of the variable with respect to which the Schwarzian is taken is now necessary.
 
By contrast,
\begin{equation}
S(\kappa(\nu),\nu)=- k_S \frac{3}{2\nu^2}
\end{equation}
is nowhere vanishing and diverges for $\nu\to 0$. However, this circumstance reveals a curious fact:
The repeated application of the Schwarzian
\begin{equation}
S(S(\kappa(\nu),\nu),\nu)=- k_S \frac{3}{2\nu^2}
\end{equation}
does not affect the result.
In other words, we have identified a fixed point of the Schwarzian as $-\frac{3 }{2x^2}$, i.e.,
\begin{equation}
S(f(x),x)=f(x) \qquad\hbox{for}\qquad f(x)=-\frac{3 }{2x^2}.
\end{equation}
This is a general property, apparently nontrivial when thinking to the associated differential equation.
Furthermore, for any real constant $a\not=0$,
\begin{equation}
S(a x^2,x)=-\frac{3 }{2x^2},
\end{equation}
i.e., $S$ attains the fixed point.
We have thus shown that the  \lq\lq Schwarzian curvature" of the family of all 
circular equatorial orbits in Schwarzschild spacetime coincides with the 
fixed point of the Schwarzian itself.

In the Kerr case the situation is quite similar. Indeed one has
\begin{equation}
\label{k_kerr}
\kappa(\nu) =k_K \frac{(\nu-\nu_1)(\nu-\nu_2)}{(1-\nu^2)},
\end{equation}
which is a fractional quadratic expression. 
Upon using the static observers 
$$
u=\frac{1}{\sqrt{-g_{tt}}}
\frac{\partial}{\partial t}\equiv u_{\rm stat}
$$ 
as fiducial observers, 
and denoting the four-velocity of the timelike equatorial circular geodesics as 
$U_\pm$ (with $U_+=U_1$, with relative velocity with respect to $u$ denoted by 
$\nu_+=\nu_1$ and $U_-=U_2$, with relative velocity with respect 
to $u$ denoted by $\nu_-=\nu_2$) one finds  
\begin{eqnarray}
  \nu_{\pm} 
    &=& \frac{\sqrt{\Delta}}{[a\pm\sqrt{r/M }(r-2M )]}
              \ ,\\ 
  \gamma(U_\pm,u) 
    &=& \left(1-\frac{2M}{r} \pm a \sqrt{\frac{M}{r^3}}\right)
       \left[\left(1-\frac{2M}{r}\right)\left(1-\frac{3M}{r}\pm 2a  
              \sqrt{\frac{M}{r^3}}\right) \right]^{-1/2}.
\end{eqnarray}
Moreover
\begin{equation}
k_K= - \frac{[r(r-2M )^2-Ma^2]}{r^2\sqrt{\Delta}(r-2M)}.
\end{equation}

The Schwarzian of $\kappa(\nu)$ (in $\nu$) is nonvanishing, and given by
\begin{equation}
S(\kappa(\nu),\nu)= -6k_K  \frac{(1-\nu_2^2)(1-\nu_1^2)}
{(\nu_1+\nu_2)^2(\nu^2-2\nu_{\rm rel}^{-1}+1)^2} 
\end{equation}
where
\begin{equation}
\nu_{\rm rel}=\frac{(\nu_1+\nu_2)}{(1+\nu_1\nu_2)}.
\end{equation}
Orbits such that $\nu^2-2\nu_{\rm rel}^{-1}+1=0$ are well known and 
correspond to extremely accelerated curves, i.e., such that
\begin{equation}
\frac{d \kappa}{d\nu}=0.
\end{equation}
We have hereby obtained a new characterization of these orbits (extremely 
accelerated, both in Schwarzschild and Kerr spacetimes, see e.g., Refs. 
\cite{Semerak:1998my,Bini:1999wn}) in terms of the 
Schwarzian of their curvature: they correspond to a diverging Schwarzian. 
This is implied by the other condition in the denominator, $\nu_1+\nu_2=0$, 
which is also satisfied by the extremely accelerated observers, known to 
\lq\lq see" the geodesics moving with equal and opposite velocities, or to 
measure zero precession when carrying a test gyroscope.
The Schwarzian vanishes instead when one or the other of the geodesics 
becomes null: $\nu_{1,2}=\pm 1$.

Interestingly, the following relation is found to be valid:
\begin{equation}
S \left(\frac{1}{\kappa(\nu)},\nu \right)=S(\kappa(\nu),\nu).
\end{equation}
Indeed, one can easily check that this is a general property of the Schwarzian:
\begin{equation}
S \left(\frac{1}{f(x)},x \right)=S(f(x),x),
\end{equation}
i.e., the Schwarzian of a function coincides with the Schwarzian of its reciprocal,
provided that $f$ is nowhere vanishing and at least of class $C^3$.

\section{Schwarzian in a spacetime model with variable mass}

There exist models of physical interest in which the mass parameter
depends linearly \cite{Berezin} on retarded time $u$ (cf. 
Ref. \cite{Ghoshal}):
\begin{equation}
M(u)=M_{0}-\mu u ,
\label{(5.1)}
\end{equation}
where $\mu$ is a positive parameter in dimensionless units.
In such a case, the surface gravity $\kappa$ depends\footnote{
In the case of ingoing null vectors $l$, one can take them in the
form ($u$ being the retarded time variable)
$$
l=\frac{\partial}{\partial u}+\frac{1}{2}
\left(1-2 \frac{M(u)}{r}\right) 
\frac{\partial}{\partial r},
$$
and hence the equation $\nabla_{l}l=- \kappa l$ implies that
$\kappa(u)=\frac{1}{4M(u)}$.} on $u$ as well, i.e.,
\begin{equation} 
\kappa(u)=\frac{1}{4M(u)}=\frac{1}{4(M_{0}-\mu u)},
\label{(5.2)}
\end{equation}
and the affine parameter $\lambda$ obeys the linear differential equation
\begin{equation}
\left[\frac{d^{2}}{du^{2}}+\kappa(u)\frac{d}{du}\right]\lambda(u)=0,
\label{(5.3)}
\end{equation}
which is the familiar equation that must be imposed in order to obtain
the affine parametrization for geodesics.
On denoting by $C_{0}$ and $C_{1}$ two
integration constants, the general solution of Eq. \eqref{(5.3)} is
\begin{equation}
\lambda(u)=C_{0}+C_{1}(M(u))^{1+\frac{1}{4\mu}}.
\label{(5.4)}
\end{equation}
The resulting Schwarzian is therefore
\begin{equation}
S(\lambda(u))=h(\mu)
\left(\frac{M'(u)}{M(u)}\right)^{2}
\label{(5.5)}
\end{equation}
and eventually, by virtue of Eq. \eqref{(5.2)},
\begin{equation}
S(\lambda(u))=h(\mu)
\left(\frac{\kappa'(u)}{\kappa(u)}\right)^{2}.
\label{(5.6)}
\end{equation}
having defined
\begin{equation}
h(\mu)=-\frac{1}{4\mu}\left(1+\frac{1}{8\mu}\right),
\label{(5.7)}
\end{equation}
This implies in turn that the squared logarithmic derivative of surface gravity 
is a projective invariant, because we can then write 
\begin{equation}
\left(\frac{\kappa'(u)}{\kappa(u)}\right)^{2}
= \frac{S(\lambda(u))}{h(\mu)}.
\label{(5.8)}
\end{equation}

\section{Further consequences for black hole physics}

The projective structure discussed in Sec. 3 determines also the following physical properties:
\vskip 0.3cm
\noindent
(i) The power $P$ radiated from a black hole, since it is proportional to the fourth 
power of the surface gravity.
\vskip 0.3cm
\noindent
(ii) The derivative of the mass content with respect to the area of the event horizon, 
as is clear from the well-established formula
\begin{equation}
\delta M=\frac{\kappa}{8\pi}\delta A+\Omega_{H}\delta J+\Phi_{H}dQ.
\label{(6.1)}
\end{equation}
Of course, in the Schwarzschild case, the second and third term on the right-hand
side of Eq. \eqref{(6.1)} do not occur.
\vskip 0.3cm
\noindent
(iii) The fundamental imaginary frequency of quasinormal modes is proportional to the 
surface gravity:
\begin{equation}
{\rm Im}(\omega_{n}) \sim -\kappa \left(n+\frac{1}{2}\right).
\label{(6.2)}
\end{equation}
Thus, eventually, the projective structure of Sec. 3 controls the decay rate of 
perturbations and the stability of the black hole.
\vskip 0.3cm
\noindent
(iv) The time necessary to return to equilibrium after a perturbation, since it is
inversely proportional to the surface gravity.

\section{Other open problems}

The framework outlined in Secs. 3, 4 and 5 is, in our opinion, of firm conceptual value,
since it suggests that black hole temperature (and hence black-hole entropy as well)
depends only on the projective structure of the event horizon. However, as far as
we can see, nothing guarantees that our projective picture survives in the presence
of horizons undergoing a fast evaporation, or for theories of gravity allowing
for causality violations, or in the presence of superluminal dispersion, or for theories of
gravity with nonvanishing torsion. We can foresee that the attempt of testing our picture
will need the investigation of several topics, e.g., time-dependent black hole
geometries, analog black holes, constraints on quantum gravity models, black hole 
thermodynamics in spacetime models which are not asymptotically flat.

Furthermore, two naturally occurring questions deserve careful consideration:
\vskip 0.3cm
\noindent
(i) Equation \eqref{(3.3)} can be used to actually define surface gravity by means of
projective methods, but does it have any relation with the differential equation
\eqref{(1.2)} for surface gravity?
\vskip 0.3cm
\noindent
(ii) Two torsion-free connections $\nabla$ and $D$ 
are said to be projectively equivalent \cite{Ov1} if there exists a $1$-form field 
$\lambda$ such that
\begin{equation}
D_{X}Y=\nabla_{X}Y+\lambda(X)Y+\lambda(Y)X.
\label{(7.1)}
\end{equation}

Can this concept be exploited in our projective approach to black hole thermodynamics?
How is surface gravity changing upon passing to a projectively equivalent connection?
As far as we can see, if we consider Eq. \eqref{(1.2)} as well as its counterpart for
the torsion-free connection $D$, i.e.,
\begin{equation}
D_{K}K=-\kappa_{D} \; K, 
\label{(7.2)}
\end{equation}
we obtain following Ref. \cite{Eastwood}, according to which ($Y_{\mu}$ being the 
components of a $1$-form)
\begin{equation}
D_{\mu}K^{\nu}=\nabla_{\mu}K^{\nu}+Y_{\mu}K^{\nu}
+Y_{\rho}K^{\rho}\delta_{\mu}^{\; \nu},
\label{(7.3)}
\end{equation}
and hence
\begin{equation}
D_{K}K^{\nu}=K^{\mu}D_{\mu}K^{\nu}
=K^{\mu}\nabla_{\mu}K^{\nu}+2(Y_{\rho}K^{\rho})K^{\nu}
=-\kappa_{D}K^{\nu}.
\label{(7.4)}
\end{equation}
In other words, Eq. \eqref{(7.2)} would imply that the vector field $K$ obeys the
nonlinear equation
\begin{equation}
K^{\mu}\nabla_{\mu}K^{\nu}=-(\kappa_{D}+2Y_{\rho}K^{\rho})K^{\nu}.
\label{(7.5)}
\end{equation}

\appendix
\section{Null hypersurfaces and Killing horizons}
 
In order to achieve self-consistency, some basic
concepts on null hypersurfaces and Killing horizons are here summarized, 
following mainly chapter 2 of Ref. \cite{PKT}.

In a spacetime manifold with metric $g$, one can consider a family of hypersurfaces 
that are level surfaces of a smooth function $\Gamma$, and hence obey the condition
$\Gamma={\rm constant}$. For each hypersurface, the vector field normal to it reads as
(on using summation over repeated indices)
\begin{equation}
l=F(x)g^{\mu \nu}(\partial_{\nu}\Gamma)\frac{\partial}
{\partial x^{\mu}},
\label{(A1)}
\end{equation}
where $F$ is a smooth nowhere vanishing function, and $g^{\mu \nu}$ are the components
of the inverse metric $g^{-1}$. 

If $g(l,l)=l^{\mu}g_{\mu \nu}l^{\nu}
=l_{\nu}l^{\nu}=0$ for a particular hypersurface $\Sigma_{N}$ in the family, such a
$\Sigma_{N}$ is said to be a {\it null hypersurface}. A vector $W$ tangent to 
$\Sigma_{N}$ is one for which $g(W,l)|_{\Sigma_N}=W_{\nu}l^{\nu}=0$. On the other hand, since
$\Sigma_{N}$ is null, the condition $g(l,l)=0$ implies that $l$ is itself a tangent 
vector, and hence one can express its components $l^{\mu}$ in the form
\begin{equation}
l^{\mu}=\frac{dx^{\mu}}{d\lambda},
\label{(A2)}
\end{equation}
for some null curve $x^{\mu}(\lambda)$ in $\Sigma_{N}$
which is the integral curve of the null vector field $l$. Moreover, the curves 
$x^{\mu}(\lambda)$ turn out to be geodesics, as is proved now, 
following Sec. 2.3.5 of Ref. \cite{PKT}. 

First, from the definition of the  
vector field $l$ in Eq. \eqref{(A1)}, one finds
\begin{equation}
g^{\mu \nu}\partial_{\nu}\Gamma=\frac{l^{\mu}}{F},
\; \partial_{\mu}\Gamma=\omega_{\mu}=\frac{l_{\mu}}{F},
\label{(A3)}
\end{equation}
while of course
\begin{equation}
l^{\rho}\frac{(\nabla_{\rho}F)}{F}=\frac{dx^{\rho}}{d\lambda}\partial_{\rho}
(\log F)=\frac{d}{d\lambda}(\log F), 
\label{(A4)}
\end{equation}
and hence, since in general relativity the covariant 
derivative of the $1$-form $\omega=d\Gamma$ is symmetric, one obtains
\begin{eqnarray}
\; & \; &
\nabla_{l}l^{\mu}=l^{\rho}\nabla_{\rho}(Fg^{\mu \nu}\partial_{\nu}\Gamma)
\nonumber \\
&=& l^{\rho}(\nabla_{\rho}F)\frac{l^{\mu}}{F}
+Fg^{\mu \nu}l^{\rho}\nabla_{\rho}\omega_{\nu}
\nonumber \\
&=& l^{\rho}(\partial_{\rho}\log F) l^{\mu}
+F l^{\rho} g^{\mu \nu} \nabla_{\nu}\omega_{\rho}
\nonumber \\
&=& l^{\mu} \frac{d}{d\lambda}(\log F)
+Fl^{\rho}\left[-F^{-2}(\nabla^{\mu}F)l_{\rho}
+F^{-1}\nabla^{\mu}l_{\rho}\right]
\nonumber \\
&=& l^{\mu}\frac{d}{d\lambda}(\log F)
+\frac{1}{2}\nabla^{\mu}l^{2}
-l^{2}(\partial^{\mu}\log F),
\label{(A5)}
\end{eqnarray}
where $l^{2}=g(l,l)$, and the identity 
$$
l^{\rho}\nabla^{\mu}l_{\rho}
=\frac{1}{2} \nabla^{\mu}(l^{2})
$$
has been exploited eventually. 
Now on a generic level surface of $\Gamma$ the
squared pseudonorm of $l$ is just constant, and hence, by covariant differentiation
along a tangent vector field $\tau$, one finds
\begin{equation}
\nabla_{\tau}l^{2}=\tau^{\mu}\nabla_{\mu}l^{2}
=\tau^{\mu}\partial_{\mu}l^{2}=0
\Longrightarrow \partial_{\mu}l^{2}=\sigma \; l_{\mu},
\label{(A6)}
\end{equation}
because tangent vector fields are orthogonal to $l$. Eventually, if
$F$ is constant, we find from Eqs. \eqref{(A5)} and \eqref{(A6)} 
that $\nabla_{l}l=\frac{\sigma}{2}l$. In particular, evaluation
of $\nabla_{l}l$ on $\Sigma_{N}$, where $g(l,l)=0$, yields
\begin{equation}
\nabla_{l}l=\left[\frac{d}{d\lambda}(\log F)+\frac{\sigma}{2}\right]l,
\label{(A7)}
\end{equation}
and hence the affine parametrization is achieved provided that
($F_{0}$ being a constant)
\begin{equation}
F=F_{0}{\rm exp}\left[-\frac{1}{2}\int_{\lambda_{0}}^{\lambda}
\sigma(w)dw \right],
\label{(A8)}
\end{equation}
where we allow for a dependence of $\sigma$ on $\lambda$, in agreement
with Eq. \eqref{(A2)}.

By definition, the null geodesics 
$x^{\mu}(\lambda)$ with affine parameter $\lambda$, for which the tangent vectors
$\frac{dx^{\mu}}{d\lambda}$ are normal to a null hypersurface $\Sigma_{N}$, are
the {\it generators} of $\Sigma_{N}$. 
A null hypersurface $\Sigma_{N}$ is said to be a {\it Killing horizon} of a Killing
vector field $K$ if, on $\Sigma_{N}$, $K$ is normal to $\Sigma_{N}$. Let $\nabla$ be
the Levi-Civita connection on spacetime. For $l$ normal to $\Sigma_{N}$ and such that
\begin{equation}
\nabla_{l}l=0
\label{(A9)}
\end{equation}
with an affine parametrization, one can exploit the proportionality relation
\begin{equation}
K=f \; l,
\label{(A10)}
\end{equation}
where $f$ is a smooth function, in order to obtain the equation
\begin{equation}
\nabla_{K}K= -\kappa \; K \; {\rm on} \; \Sigma_{N},
\label{(A11)}
\end{equation}
where
\begin{equation}
\kappa=-k^{\mu} \partial_{\mu} \log |f|
\label{(A12)}
\end{equation}
is the surface gravity. We here follow the sign convention of Ref. \cite{H2}
for $\kappa$, and hence we change sign with respect to Ref. \cite{PKT}.
In the particular case of Schwarzschild spacetime, the concepts of Killing
and event horizon coincide.

On the Killing horizon $\Sigma_{N}$ one can choose coordinates such that
\begin{equation}
K=\frac{\partial}{\partial u},
\label{(A13)}
\end{equation}
except at points where the Killing vector field vanishes. If $u$ is a function
of $\lambda$ on an orbit of $K$ with affine parameter $\lambda$, one has
\begin{equation}
\left . K \right |_{\rm orbit}
=\frac{d\lambda}{du}\frac{d}{d\lambda}=fl,
\label{(A14)}
\end{equation}
where 
\begin{equation}
f=\frac{d\lambda}{du},
\label{(A15)}
\end{equation}
and (see Eq. \eqref{(A2)})
\begin{equation}
l=\frac{d}{d\lambda}=\frac{dx^{\mu}(\lambda)}{d\lambda}\partial_{\mu}.
\label{(A16)}
\end{equation}
By virtue of Eqs. \eqref{(A6)} and \eqref{(A7)} one has
\begin{equation}
\kappa=-\frac{\partial}{\partial u}\log |f|,
\label{(A17)}
\end{equation}
which is constant on orbits of $\Sigma_{N}$. For such orbits, 
$f=f_{0}e^{-\kappa u}$ for an arbitrary constant $f_{0}$. Upon exploiting the
freedom to shift $u$ by an arbitrary constant, one can choose $f_{0}=\pm \kappa$
and hence, up to an additive constant for the affine parameter $\lambda$, one has
\begin{equation}
\frac{d\lambda}{du}=\pm \kappa e^{-\kappa u} \Longrightarrow 
\lambda = \pm e^{-\kappa u}.
\label{(A18)}
\end{equation}
As $u$ ranges from $-\infty$ to $+ \infty$, the two portions ($\lambda >0$ or
$\lambda <0$) of the generator of $\Sigma_{N}$ are covered. At $\lambda=0$ a fixed
point of $K$ occurs, which is said to be the bifurcation $2$-sphere.

\section{Projective differential geometry}

In this appendix we rely mainly upon the work in Ref. \cite{Lane1} in order to provide
physics-oriented readers with a pedagogical review of some basic properties of 
projective differential geometry (while the more advanced modern literature has
been already cited at the beginning of our paper).

As is well known, metric geometry is the study of those properties of figures which
are invariant under the group of rigid motions. Examples of metric invariants are
the distance between two points, the angle between two lines, area, shape, size. 
On the other hand, projective geometry studies those properties of figures that
are invariant under the group of projections. An examples is provided by the
straightness of a line: if a line is straight before projection, it remains straight 
afterwards. Another example is the united position of point and line: if a point 
is on a line before projections, it will be on the line afterwards. Other projective
invariants are the cross-ratio of four points on a line, the harmonic separation
of four collinear points or of four coplanar concurrent lines. 

Eventually, projective differential geometry is both projective and differential. 
Since the group of projections contains as a subgroup the group of rigid motions,
and since every invariant under a group is also an invariant under a subgroup of
that group, it follows that all of projective geometry may properly be included
in metric geometry, and that in particular all of projective differential 
geometry may be included in metric differential geometry. However, the task remains
of selecting from metric differential geometry the concepts that are actually 
projective. For example, the metric normal at a point of a surface is replaced by 
the projective normal \cite{T2}.

The formulation of projective differential geometry by means of differential forms 
is in our opinion particularly well suited for introducing this branch of mathematics to
theoretical physicists. Fubini \cite{F1,F2,F3,F4} studied the problem of defining
a surface by differential forms, except for a projective transformation in ordinary
space, and for this purpose he considered three differential forms, two quadratic
and one cubic, which can be written in the form
\begin{equation}
\beta \gamma (du \otimes dv + dv \otimes du), \;
2 \beta \gamma (\beta du \otimes du \otimes du
+\gamma dv \otimes dv \otimes dv), \;
p du \otimes du -q dv \otimes dv.
\label{(B1)}
\end{equation}
Fubini discovered that $\beta,\gamma,p,q$ must be functions of $u$ and $v$ that, 
after defining (the subscript denoting differentiation with respect to the variable)
\begin{equation}
\theta=\log(\beta \gamma), \;
\phi=(\log \beta \gamma^{2})_{u} , \;
\psi=(\log \beta^{2}\gamma)_{v},
\label{(B2)}
\end{equation}
obey the system of differential equations 
\begin{equation}
\theta_{uvv}=(\gamma \phi)_{u}+2 q_{u}+\theta_{v}\theta_{uv}-\beta \gamma \psi,
\label{(B3)}
\end{equation}
\begin{equation}
\theta_{uuv}=(\beta \psi)_{v}+2p_{v}+\theta_{u}\theta_{uv}-\beta \gamma \phi,
\label{(B4)}
\end{equation}
\begin{equation}
p_{vv}-\theta_{v}p_{v}+\beta q_{v}+2q \beta_{v}
=q_{uu}-\theta_{u}q_{u}+\gamma p_{u}+2p \gamma_{u},
\label{(B5)}
\end{equation}
which turn out to be the integrability conditions for the system of equations
provided by the differential-equations approach to the theory of surfaces.
The differential forms \eqref{(B1)} determine a non-ruled surface referred to its 
asymptotic curves\footnote{For which the normal curvature vanishes, which implies the 
vanishing of the extrinsic curvature tensor when evaluated on vectors tangent to 
the surface.} in ordinary space, except for a projective transformation. 

As we mentioned before, the normal line at a point of a surface cannot be defined in
the projective theory, but Green and Fubini discovered the projective normal
\cite{Lane1,Lane2,T2}. The geodesic curves of metric theory do not have a projective
character, but they have stimulated various investigations and generalizations in
projective theory, including also the so-called pangeodesics \cite{Segre}. Moreover, 
in metric theory the parametric net is said to be isothermally orthogonal if the
coefficients of the first fundamental form of surfaces:
\begin{equation}
E du \otimes du + F (du \otimes dv + dv \otimes du)
+G (dv \otimes dv)
\label{(B6)}
\end{equation}
satisfy the equations
\begin{equation}
F=0, \; (\log E G^{-1})_{uv}=0.
\label{(B7)}
\end{equation}
The counterpart in projective differential geometry is the concept of isothermally
conjugate net, for which the coefficients of the second fundamental form
(i.e., the extrinsic curvature tensor)
\begin{equation}
L du \otimes du + M (du \otimes dv + dv \otimes du)
+N (dv \otimes dv)
\label{(B6)}
\end{equation}
satisfy the equations
\begin{equation}
M=0, \; (\log EN^{-1})_{uv}=0.
\label{(B9)}
\end{equation} 

In classical differential geometry the problem of (local) isometric embedding of a
two-dimensional surface into three-dimensional Euclidean or Riemannian space has
led to dedicated efforts for more than a century, and the original nomenclature
used to call it {\it applicability theory}. In projective differential geometry,
the {\it projective applicability of two surfaces} can be defined as follows
\cite{Villa}. A bijective correspondence among the points of two surfaces
$\Sigma,{\overline \Sigma}$ is said to be a projective applicability if, for each
point $A$ of $\Sigma$ one can find a homography $\Omega$ that carries $A$ into
the corresponding point ${\overline A}$ of ${\overline \Sigma}$ and carries 
a curve ${\cal C}$, coming out of $A$ and belonging to $\Sigma$, into a curve
${\cal C}_{\Omega}$ which has at ${\overline A}$ a second-order 
contact\footnote{This means that derivatives of ${\overline {\cal C}}$ and ${\cal C}_{\Omega}$ 
of order $0,1,2$ are equal at ${\overline A}$.}
with the curve ${\overline {\cal C}}$ (coming out of ${\overline A}$) corresponding 
to the curve ${\cal C}$.

The {\it projective line element} of a surface, introduced by Fubini, is the counterpart
of the line element of metric geometry. Given a surface $\Sigma$, let us consider the 
two asymptotic tangents \cite{Villa} coming out of a point $O$, and those coming out
another point $O'$ of $\Sigma$. These four lines determine four points on the line 
obtained by intersection of the planes tangent at $O$ and $O'$. The cross-ratio of 
these four points, assuming $O'$ infinitesimally close to $O$, is equal to 
$\frac{4}{9}\Phi^{2}$, where $\Phi$ is then said to be the projective line element 
of the surface \cite{F2,F3,T1,Villa}. Remarkably, if $u$ and $v$ are asymptotic coordinates
on the surface, the projective line element $\Phi$ can be expressed as the ratio
(up to a multiplicative constant) of two of the differential forms in Eq. \eqref{(B1)}: 
\begin{equation}
\Phi=\frac{(\beta du \otimes du \otimes du + \gamma dv \otimes dv \otimes dv)}
{(du \otimes dv+dv \otimes du)}.
\label{(B10)}
\end{equation}
This notation is standard in the old literature on projective differential geometry,
but deserves a last effort in defining its meaning. Equation \eqref{(B10)} means that
we consider a map $\cal F$ which, out of the first two differential forms in Eq.
\eqref{(B1)}, engenders a projectively invariant geometric object $\Phi$, 
the projective line element of Fubini. The multiplicative factors $\beta \gamma$ 
of Eq. \eqref{(B1)} have
been factored out, exactly as it would occur in a ratio of functions.
At a deeper mathematical level, $\Phi$ is the section of a projective line bundle.
The 3-form in \eqref{(B10)} is a density of weight 1, the 2-form in \eqref{(B10)} 
is a density of weight 0, so that
$\Phi$ has weight 1. The necessary and sufficient condition for two surfaces to be 
projectively applicable is the equality of their projective line elements.

Yet other results of classical differential geometry have their projective counterpart.
For example, a theorem of Meusnier, according to which the osculating circles of all
plane sections at a point of a surface lie on a sphere, has a counterpart in a theorem
by Moutard, according to which the osculating conics of all plane sections at a point
of a surface lie on a quadric surface \cite{Lane1}.

\section*{Acknowledgements}
D.B. and G.E. are grateful to INDAM for membership. G.E. thanks the ET and QGSKY  
research units of INFN Naples, and Yannick Herfray for invitation to Tours, where
this research was initiated, and for scientific conversations.

\end{document}